\documentclass[12pt]{iopart}
\usepackage{graphics}
\usepackage{graphicx}

\usepackage{iopams}
\begin{document}
\title[Determination of the spatial TDR-sensor characteristics...]
{Determination of the spatial TDR-sensor characteristics in strong dispersive subsoil using 3D-FEM
frequency domain simulations in combination with microwave dielectric spectroscopy}

\author{Norman Wagner, Eberhard Trinks, Klaus Kupfer}

\address{Institute of Material Research and Testing  at the Bauhaus-University Weimar,\\ Amalienstr. 13, 99423 Weimar
        \\Tel: ++49-3643-564-221\\Fax: ++49-3643-564-202} \ead{norman.wagner@mfpa.de}
\begin{abstract}
The spatial sensor characteristics of a 6cm TDR flat band cable sensor section was simulated with
finite element modelling (High Frequency Structure Simulator-HFSS) under certain conditions: (i) in
direct contact to the surrounding material (air, water of different salinities, different synthetic
and natural soils (sand-silt-clay mixtures)), (ii) with consideration of a defined gap of different
size filled with air or water and (iii) the cable sensor pressed at a borehole-wall. The complex
dielectric permittivity $\varepsilon^\star(\omega, \tau_i)$  or complex electrical conductivity
$\sigma^\star(\omega, \tau_i)=i\omega\varepsilon^\star(\omega, \tau_i)$  of the investigated
saturated and unsaturated soils was examined in the frequency range 50MHz-20GHz at room temperature
and atmospheric pressure with a HP8720D- network analyser. Three soil-specific relaxation processes
are assumed to act in the investigated frequency-temperature-pressure range: one primary
$\alpha$-process (main water relaxation) and two secondary ($\alpha'$, $\beta$)-processes due to
clay-water-ion interactions (bound water relaxation and the Maxwell-Wagner effect). The dielectric
relaxation behaviour of every process is described with the use of a simple fractional relaxation
model. 3D finite element simulation is performed with a $\lambda/3$ based adaptive mesh refinement
at a solution frequency of 1MHz, 10MHz, 0.1GHz, 1GHz and 12.5GHz. The electromagnetic field
distribution, S-parameter and step responses were examined. The simulation adequately reproduces
the spatial and temporal electrical and magnetic field distribution.  High-lossy soils cause
depending on increasing gravimetric water content and bulk density an increase of TDR signal rise
time as well as a strong absorption of multiple reflections. Air or water gap work as quasi
wave-guide, i.e. the influence by surrounding medium is strongly reduced. Appropriate
TDR-travel-time distortions can be quantified.

\end{abstract}

\maketitle

\section{Introduction}

Soil science, geophysical prospecting, agriculture, hydrology, archeology and geotechnical
engineering have benefited greatly from developments in radio and microwave technology.
Electromagnetic techniques are used to estimate soil and rock physical characteristics such as
water content, density and porosity (\cite{Shen85}, \cite{Olhe81}, \cite{Kupf05}, \cite{Kell05}).
Both invasive methods, such as time domain reflectometry (\cite{Topp80}, \cite{Robi03},
\cite{Heimo04}) and cross borehole radar \cite{Fech04}, and noninvasive methods, such as capacity
methods (\cite{Seyf04}, \cite{Kell05}, \cite{Kupf05}) and ground penetrating radar (\cite{Kats74},
\cite{Anna75}, \cite{Ross75}, \cite{Davi89}, \cite{Hana06}) are used. Common to all these
techniques is the fact that electromagnetic wave interaction depends on dielectric properties of
rock or soil deposit through which it travels, which are influenced by chemical composition,
mineralogy, structure, porosity, geological age and forming conditions. Besides, several additions
like ubiquitous water have an effect on the dielectric properties.

In particular, knowledge of the spatial and temporal variability of water saturation in soils is
important to obtain improved estimates of water flow (and its dissolved components) through the
vadose zone. Due to its accuracy and potential for automated measurement, TDR has become one of the
standard methods to measure spatial and temporal variability of water contents in laboratory soil
cores and experimental field plots \cite{Heimo04}. For this purpose, the object of numerous
experimental and theoretical investigations is the development of general dielectric mixing models
for a broad class of soil textures and structures (\cite{Topp80}, \cite{Roth92}, \cite{Topp02}).
Mostly, these empirical, numerical or theoretical models base on the assumption of a constant
dielectric permittivity of the soil as a function of volumetric water content in a narrow frequency
range around 1GHz (\cite{Saar98}, \cite{Sihv00}, \cite{Cose04}, \cite{Evet05}, \cite{Rega06}).
However, the strong frequency dependence in the dielectric relaxation behaviour below 1GHz due to a
certain amount of swelling clay minerals in nearly each real soil is considered only insufficiently
(\cite{Logs2004}, \cite{Heimo04}, \cite{Logs2005}, \cite{Kell05}).

The type of multi-scale structure renders the analysis of dynamic data in clays rather complex. The
problem has been addressed both by experimental and modelling techniques. Besides broadband
dielectric spectroscopy on clay-water suspensions microscopic simulations of clays have been an
active field of research since the late 1980s and began with simulations at ambient temperature and
pressure, of clays with various cationic species. More recently, several studies appeared dealing
with non-ambient conditions (increased temperatures and pressures), which are primarily linked to
the issue of storage of radioactive waste or bore-hole stability \cite{Mali05}. Previous
experimental and modelling results suggest that clay-water systems have multiple relaxation
processes, such as interfacial polarizations around the clay particles and rotational relaxation of
bound and free $H_2O$. Therefore, the dielectric behavior is expected to be complicated. Useful and
precise dielectric information may only be obtained when each relaxation process is extracted from
the complicated overall behavior based on the measurement of the complex dielectric permittivity
over a broad frequency range and at high resolutions (\cite{Ishi00}, \cite{Ishi03}, \cite{Logs04},
\cite{Logs05}).

In the present study, the the dielectric relaxation behaviour of the investigated saturated and
unsaturated soils was examined in the frequency range 50MHz-20GHz. To parameterize the dielectric
spectra three soil-specific relaxation processes are assumed to act as a function of water
saturation and porosity in the investigated frequency-temperature-pressure range: one primary
$\alpha$-process (main water relaxation) and two secondary ($\alpha'$, $\beta$)-processes due to
clay-water-ion interactions (bound water relaxation and the Maxwell-Wagner effect). The dielectric
relaxation behaviour of every process is described with the use of a simple fractional relaxation
model (\cite{Ishi00}, \cite{Ishi03}, \cite{Jons77}, \cite{Holl98}, \cite{Hilf02}). The chosen
approach enables a characterization of the dielectric relaxation behaviour with the separation in
the observed relaxation processes in dependence of the porosity and the water content.

A simplification frequently utilized in TDR applications is the use of an idealized equivalent
circuit for the sensor without consideration of losses due to the skin-effect or radiation from
the sensor as well as the assumption of a homogeneous sensitivity distribution along the sensor
(\cite{Heimo04}, \cite{Hueb05}, \cite{Schl05}, \cite{Leid05}). In addition, a frequently arising
problem in various applications is the direct contact between sensor, e.g. a flexible band cable,
and surrounding medium. An air or water gap between sensor and soil leads to dramatic under or
overestimation of water content. A suitable tool for an examination of this specific problems
offers three-dimensional numeric finite element simulation (\cite{Leid05}). For these reasons, in
this study the spatial sensor characteristics of a 6cm TDR flat band cable sensor section was
simulated with electromagnetic finite element modelling (Ansoft-HFSS$^{TM}$ , High Frequency
Structure Simulator). In order to carry out the finite element calculations as realistically as
possible, the measured frequency-dependent dielectric permittivity was considered. Moreover, the
simulations were performed under certain conditions: (i) in direct contact to surrounding
material, (ii) with consideration of a defined gap of variable size filled with air or water and
(iii) cable sensor pressed at a borehole-wall.

\section{Theoretical Background}

\begin{figure}[t]
\begin{indented}
 \item[]
  \includegraphics[scale=2.48]{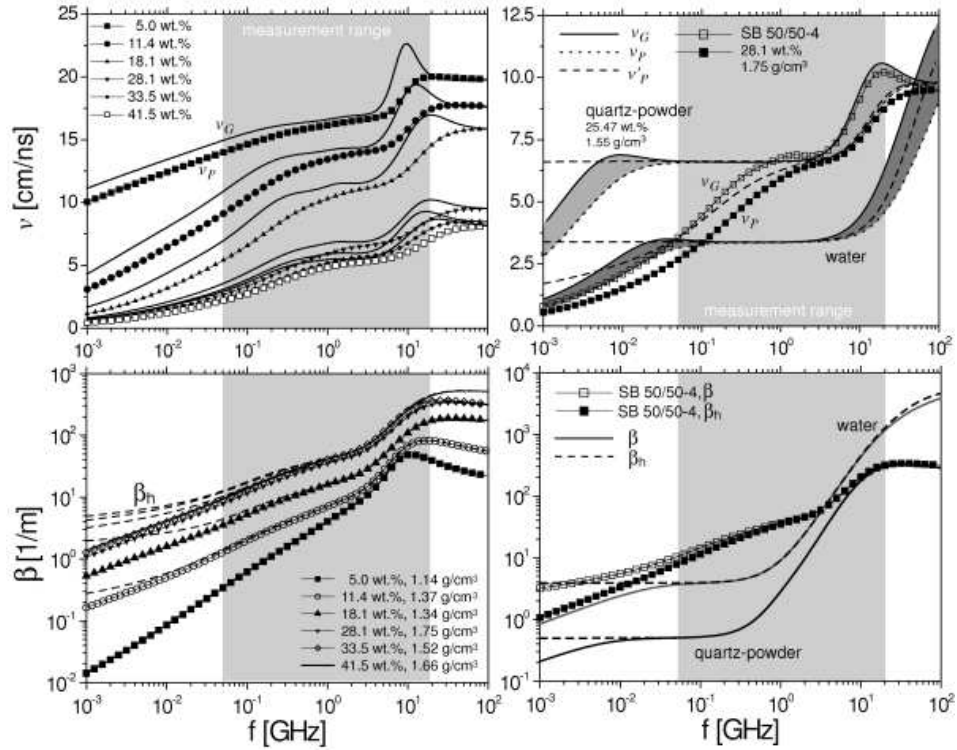}\\
  \caption{(top) Phase velocity $v_P$, high frequency approximation $v_p'$, corresponding group velocity $v_g$
   as well as (bottom) attenuation $\beta$ and corresponding high frequency approximation
   $\beta_h$ of (left) a sand-bentonite mixture (SB 50/50 1-6) with different gravimetric water content $\theta_g$ and bulk density $\varrho$
     (see section \ref{sec:measure}, Tab. \ref{tab:fit_data}, Fig. \ref{fig:complex_permittivity_bentonite}).
     (right) Comparison of natural water with $\sigma_{DC}=0.18S/m$, a mikrosil quartz powder
     and SB 50/50-4 (see Tab. \ref{tab:Calcigel} and
      \ref{tab:fit_data} below). The calculations were carried out under utilization of the measured dielectric permittivity.}\label{fig:Dispersion}
\end{indented}
\end{figure}

Time domain reflectometry measures the propagation velocity of a broadband step voltage pulse
(typical values: rise time $t_r\approx70ps$, sampling increment $\Delta t\approx20ps$) with a
bandwidth of around 20kHz to ~25GHz (Nyquist-frequency: $f_{max}=0.5/\Delta t$). But due to the
limitations of used connectors, type of the TDR device, coaxial cable type and length the effective
bandwidth is reduced distinctly \cite{Logs00}. Under atmospheric conditions the velocity of this
signal is a function of the frequency $\omega=2\pi f$ and temperature $T$ dependent effective
relative complex permittivity $\varepsilon_{{\mbox{eff}}}^\star(\omega,
T)=\varepsilon_{\mbox{eff}}'(\omega, T)-j\varepsilon_{\mbox{eff}}''(\omega, T)$ of the material
through which it travels. The overall losses $\varepsilon_{\mbox{eff}}''(\omega,
T)=\varepsilon_d''(\omega, T)+\frac{\sigma_{DC}(T)}{\omega\varepsilon_0}$ of the material which
have to be considered result from the dielectric losses $\varepsilon_d''(\omega, T)$ and the
conductive losses $\frac{\sigma_{DC}(T)}{\omega\varepsilon_0}$ due to a direct current electrical
conductivity $\sigma_{DC}(T)$. Here, $\varepsilon_0$ is the permittivity of free space and $j^2=-1$
is the imaginary unit. It is often convenient to consider the analogy of propagation phase velocity
and attenuation of an electromagnetic plane wave (see \cite{Topp80}, \cite{Davi89}, \cite{Heimo04},
\cite{Robi05}):
\begin{equation}\label{eq:velocity}
v_P (\omega, T) = c\sqrt 2 \left( {\sqrt {\sqrt {\varepsilon'_{{\mbox{eff}}} (\omega, T)^2 +
\varepsilon''_{{\mbox{eff}}} (\omega, T)^2}  + \varepsilon'_{{\mbox{eff}}} (\omega, T)} } \right)^{
- 1}
\end{equation}
\begin{equation}\label{eq:beta}
\beta (\omega, T) = \omega \sqrt {\sqrt {\varepsilon '_{{\mbox{eff}}}(\omega, T)^2  + \varepsilon
''_{\mbox{eff}}(\omega, T)^2 } - \varepsilon '_{\mbox{eff}}(\omega, T) }  \cdot \left( {c\sqrt 2 }
\right)^{ - 1}
\end{equation}
where $c=\sqrt{\varepsilon_0\mu_0}^{-1}$ is the velocity of light and $\mu_0$ the magnetic
permeability of vacuum. Hence, any modulation of an electromagnetic wave in a real medium will
propagate at a group velocity according to the Rayleigh equation (\cite{Fork89}, \cite{Grei98}):
\begin{equation}
v_g  = \frac{{d\omega }}{{dk}} = v_P \left[ {1 - \frac{f}{{v_P }}\frac{{dv_P }}{{df}}}\right]^{-1}
\end{equation}
herein, $k$ denotes the wave number. The flat band cable of length $l$ consists of three strip
conductors embedded in a polyethylene band (see \cite{Hueb05}, \cite{Kupf05} for details). The
effective group or phase velocity of the signal $v'_P$ in a perfect dielectric (pure real
dielectric constant $\varepsilon_r=\varepsilon_{\mbox{eff}}^\star=\hbox{const}$ without dispersion
and conducting losses) surrounding the cable sensor is in principle only a crude approximation of
the real material properties especially at frequencies $f<1GHz$ and $f>10GHz$ (c.f. Fig.
\ref{fig:Dispersion})
\begin{equation}\label{eq:epsilon_trawel_time}
v'_P  = v_g  = \frac{{2l}}{t} = \frac{c}{{\sqrt {\varepsilon _r } }}
\end{equation}
where $t$  is two way travel time. Considering anomalous dispersion equation
(\ref{eq:epsilon_trawel_time}) is referred to as a high frequency approximation of  phase velocity
(Fig. \ref{fig:Dispersion}). In contrast the high frequency attenuation approximation $\beta _h
\left( \omega, T\right)$ for real soils frequently used in ground penetrating radar applications
works considerably well (c.f. \cite{Davi89}, Fig. \ref{fig:Dispersion})
\begin{equation}
\beta _h \left( \omega, T\right) = \frac{{\omega\varepsilon _0 \varepsilon
''_{{\mbox{eff}}}(\omega, T) }}{{\sqrt {\varepsilon '_{{\mbox{eff}}} (\omega, T)} }}\frac{Z_0}{2}
\end{equation}
with impedance $Z_0=c\mu_0$ of vacuum. We now consider the soil as a four-phase medium composed of:
air, quartz grain, water and clay. In the particular case of spatial TDR the surrounding medium in
the direction oft the band cable is described by a relative effective permittivity
$\varepsilon^\star_{\mbox{eff}}(x, \omega, T, p)$. It depends on position $x$, angular frequency
$\omega$ and contribution due to several relaxation processes via relaxation time $\tau(T,p)$ on
absolute temperature $T$ and pressure $p$ (\cite{Ishi00}, \cite{Logs2004})
\begin{equation}
\tau _i (T,p) = \kappa _i \frac{h}{{k_B T}}\exp \left( {\frac{{E_{a,i} (T,p)}}{{RT}}} \right)
\end{equation}
Herein, $h$ denotes the Planck-constant, $k_B$-Boltzmann constant, $\kappa_i\approx 1$ the
transmission coeffitient, $R$ gas constant and $E_{a,i} (T,p) = \Delta G_i (T,p) + T\Delta S_i
(T,p)$ activation energy with free enthalpy $\Delta G_i(T,p)$ and activation entropy $\Delta
S_i(T,p)$ of the $i$-th process (\cite{Ishi00}, \cite{Ishi03}). Dielectric loss spectra of
saturated and unsaturated soils very often show a marked deviation from simple Debye-behaviour
(\cite{Hoek74}, \cite{Holl98}, \cite{Ishi00}, \cite{Kell05}). Based on the theory of fractional
time evolutions Hilfer \cite{Hilf02} derived a Jonscher type function \cite{Jons77} for the complex
frequency dependent dielectric permittivity of amorphous and glassy materials
\begin{equation}\label{eq:relaxation_model}
\tilde \varepsilon _{{\mbox{eff}},i} (\omega ,\tau _i ) - \varepsilon _\infty   = \frac{{\Delta
\varepsilon _i(T) }}{{\left( {j\omega \tau _i } \right)^{\alpha _i }  + \left( {j\omega \tau _i }
\right)^{\beta _i } }}
\end{equation}
with high frequency limit of permittivity $\varepsilon_\infty$, relaxation strength
$\Delta\varepsilon_i(T)$ as a function of temperature, angular frequency $\omega$ and stretching
exponents $0\leq\alpha_i, \beta_i$ similar to the familiar empirical Havriliak and Negami
\cite{Havr67}, Cole-Cole \cite{Cole41}, Cole-Davidson \cite{Davi51} or Kohlrausch-Williams-Watts
(\cite{Kohl47}, \cite{Will70}) dispersion and absorption functions. For the particular case
$\alpha_i=0$ and $\beta_i=1$ (\ref{eq:relaxation_model}) transforms to the Debye model.

\section{Material and Experiments}\label{sec:measure}
\begin{figure}[t]
\begin{indented}
 \item[]
  \includegraphics[scale=2.45]{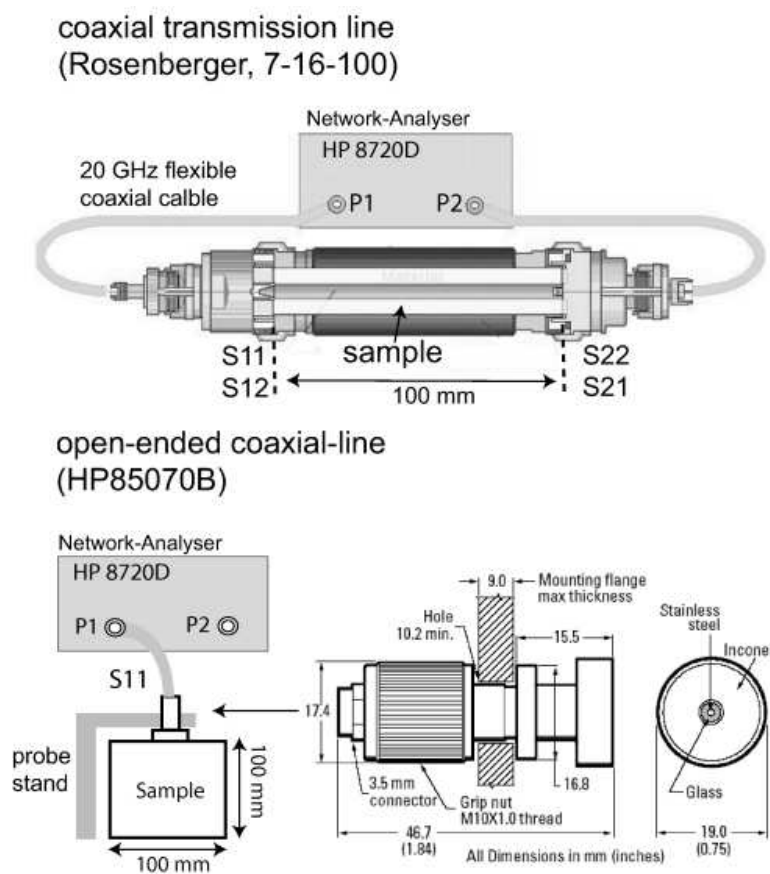}
  \caption{Schematic diagram of the experimental set-up for measuring the dielectric permittivity: (top)
  longitudinal section of the coaxial transmission line with an outer diameter of the inner conductor $d_i=7 mm$, an inner diameter of
  the outer conductor $d_o=16 mm$ and a length $l=100mm$, (bottom) open-ended coaxial-line (HP85070B).}\label{fig:measure}
\end{indented}
\end{figure}
The complex effective dielectric permittivity $\varepsilon_{\mbox{eff}}^\star$ of saturated and
unsaturated soils was examined in the frequency range 50MHz-20GHz at room temperature and
atmospheric pressure with a HP8720D- network analyser. This was performed using a combination of
open-ended coaxial-line (HP85070B) and coaxial transmission line technique (sample holder
(7x16x100)mm$^3$) (Fig. \ref{fig:measure}). The synthetic soil samples were incrementally wetted
from air dry up to saturation with natural water and equilibrated 12h. From the prepared sample a
subsample was taken with a retaining ring (hight 100 mm, inner diameter 100 mm). The retaining ring
was used as the sample holder for the HP85070B probe. Care was taken to pack the soil in the
transmission line to a  homogeneous bulk density $\varrho$ and to a constant volume. After each
dielectric measurement bulk density $\varrho$ as well as gravimetric water content $\theta_g$ were
determined. The measured complex S-parameter values $S_{ij}$ were used to calculate complex
dielectric permittivity with a commercial software (HP 85070/71C Materials Measurement Software)
after calibration with open, short, and load standards.

Different natural and synthetic soils were investigated. Here, we present our results for synthetic
soil SB50/50. It is a mixture of 50wt.\% sand (grain size $<$2mm) and 50 wt.\% bentonite (Calcigel:
71wt.\% Ca- dioctahedral smectite, 9wt.\% illite/dioctahedral mica, 1wt.\%kaoline, 1wt.\% chlorite,
9wt.\% quartz, 5wt.\% feldspar, 2wt.\% calcite, 2wt.\% dolomite).

\begin{figure}[t]
\begin{indented}
 \item[]
  \includegraphics[scale=2.3]{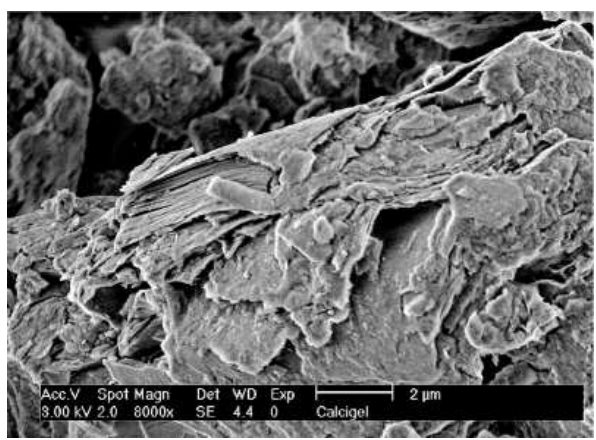}\label{fig:SEM-EDX}
  \caption{SEM-EDX micrograph of the investigated bentonite Calcigel, Philips XL30 ESEM-FEG.}
\end{indented}
\end{figure}

Ca- dioctahedral smectites are clay minerals, i.e. they consist of individual crystallites the
majority of which are $<$2$\mu$m in largest dimension. The crystal structure of clays has been
established from X-ray diffraction studies for almost all types of common clays. Smectite
crystallites themselves are three-layer clay minerals. Individual clay layers consist of fused
sheets of octahedra of Al$^{3+}$ or Mg$^{2+}$ oxides and tetrahedra of Si$^{4+}$ oxides.
Substitution of Al$^{3+}$, Mg$^{2+}$ or Si$^{4+}$ with lower valence ions results in an overall
negative charge on the layers, which is then compensated by cationic species (counterions, in the
current case $Ca^{2+}$) between clay layers (interlayers) (\cite{Lieb85}, Fig. \ref{fig:SEM-EDX}).
\begin{table}[t]
  \centering
  \caption{Physical properties of the investigated bentonite ($^\flat$\cite{Enge03}) in
  comparison to the commercial Mikrosil quartz powder (type 350, median grain size 11$\mu m$).}\label{tab:Calcigel}
  \begin{tabular}{|l|r|r|}
    \hline & Ca-Bentonite & Mikrosil\\
    \hline
    Dry density  $\varrho_d$  & 1.33 g/cm$^3$ & 0.69 g/cm$^3$\\
    Specific surface area & 493 m$^2$/g$^{(\flat)}$ & 0.38 m$^2$/g \\
    Density & 2.847 g/cm$^3$ & 2.65 g/cm$^3$ \\
    Cation exchange capacity & 62.0 meq/100g$^{(\flat)}$ &\\\hline
  \end{tabular}
\end{table}

Under increasing relative humidity the compensating ions become hydrated and the spacing between
individual layers increases. While the detailed swelling characteristics of a smectite (interlayer
spacing as a function of relative humidity) depends crucially on the charge of the clay layers
(magnitude and localisation) and the nature of the compensating ion, in general, it occurs in three
stages. Swelling begins in a step-wise manner (discrete layers of water formed in the interlayer,
states referred to as monolayer/monohydrated, bilayer/ bihydrated, etc.), becomes continuous
thereafter and in the extreme a colloidal suspension of clay particles ($<$10 aligned layers) is
formed. Beyond the microscopic scale, aggregates of aligned clay layers form particles of the order
of 10 nm - 1000 nm in size, with porosities on the meso- (8 nm - 60 nm) and macroscale ($>$60 nm)
(\cite{Mali05} and citations in it). In Tab. \ref{tab:Calcigel} the physical and chemical
properties of the used bentonite calcigel are summarized.
\begin{figure}[t]
\begin{indented}
 \item[]
  \includegraphics[scale=2.5]{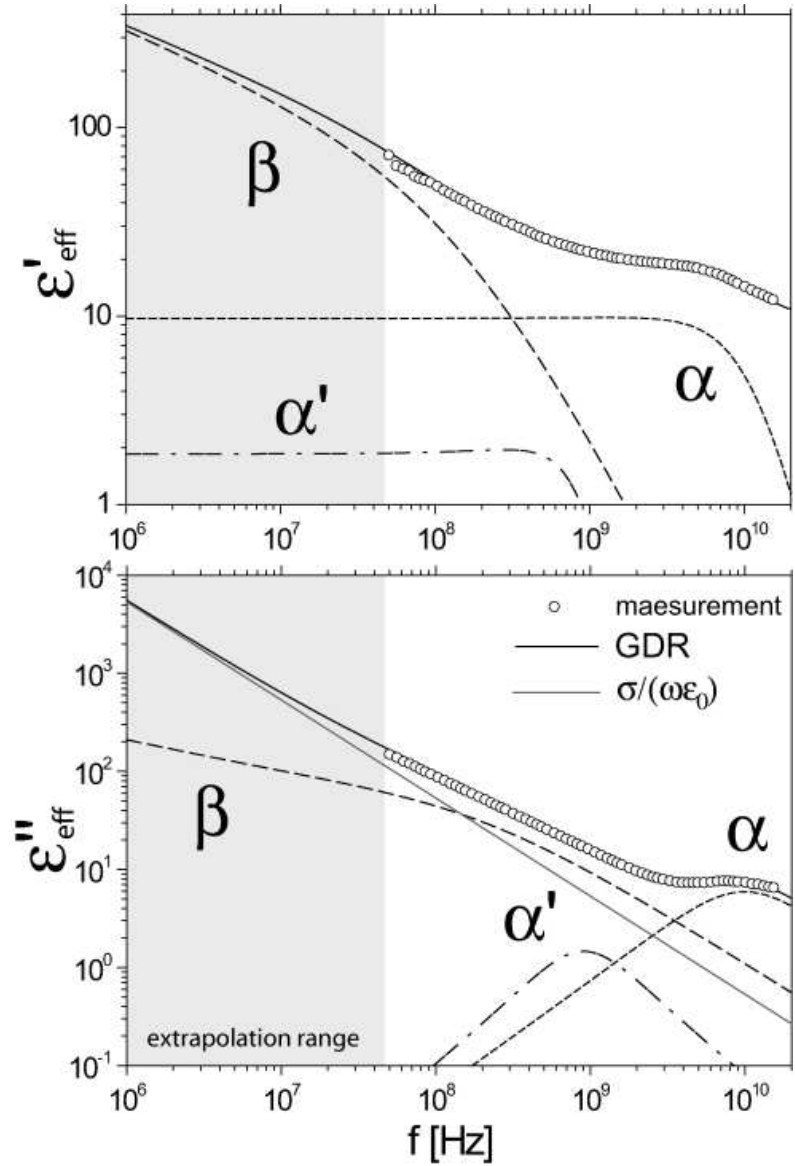}
  \caption{Complex relative dielectric permittivity $\varepsilon^\star_{\mbox{eff}}=\varepsilon'_{\mbox{eff}}-j\varepsilon''_{\mbox{eff}}$ as a function of frequency
  of sand-bentonite-mixture SB50/50-4 with GDR-fit under consideration of three relaxation processes: one primary $\alpha$-process (main water relaxation) and two secondary ($\alpha',
\beta$)-processes due to clay-water-ion interactions (bound water relaxation and the Maxwell-Wagner
effect). In this way it was possible to  separate the singe relaxation processes and to extrapolate
the dielectric permittivity to 1MHz. }\label{fig:GDRFit}
\end{indented}
\end{figure}

\begin{figure}[t]
\begin{indented}
 \item[]
  \includegraphics[scale=2.55]{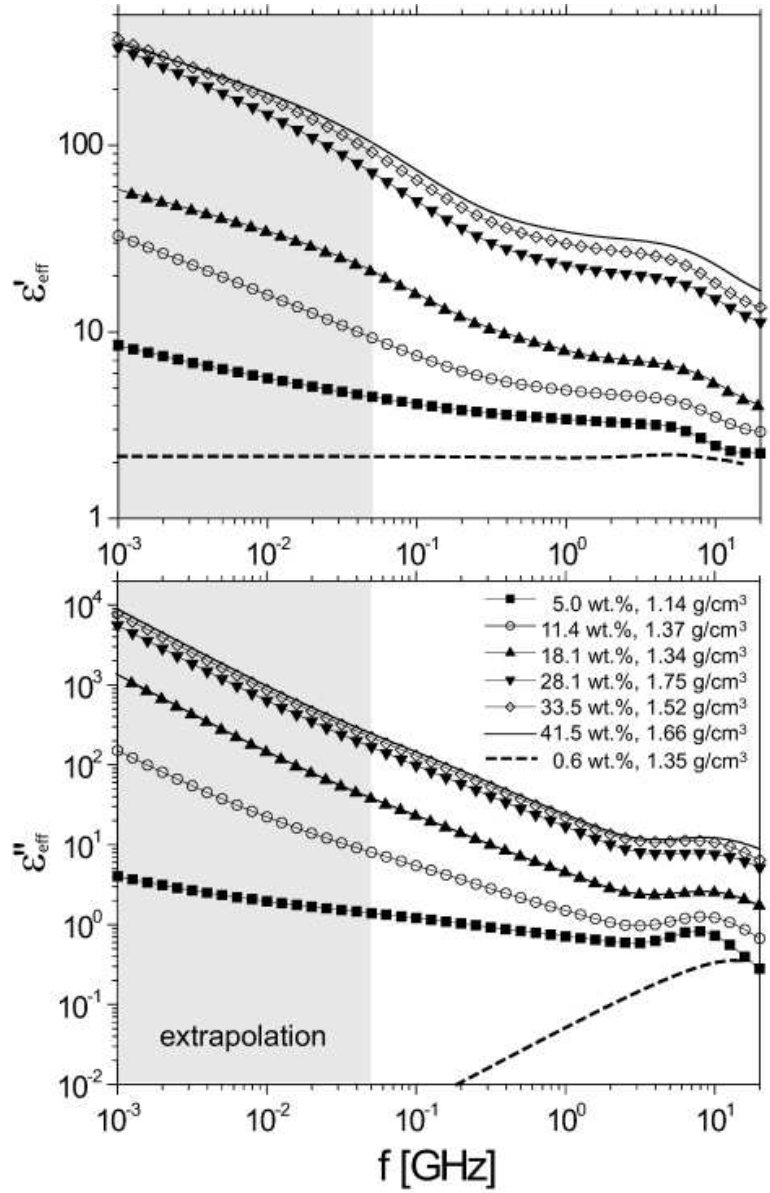}
  \caption{Dispersion and absorption curve of sand-bentonite-mixture (SB50/50) for
  seven gravimetric water contents $\theta_g$ and bulk densities $\varrho$.}\label{fig:complex_permittivity_bentonite}
\end{indented}
\end{figure}

Three relaxation processes are assumed to act in the investigated frequency-temperature-pressure
range: one primary $\alpha$-process (main water relaxation) and two secondary ($\alpha',
\beta$)-processes due to clay-water-ion interactions (bound water relaxation and the
Maxwell-Wagner effect, see Fig. \ref{fig:GDRFit}). The effective permittivity of a multiphase
soil-mixture $\varepsilon^\star_{\mbox{eff}}$ can be determined by the complex relative
permittivity of water $\varepsilon^\star_w$, bound water $\varepsilon^\star_{bw}$, the
contribution due to clay-water-ion interaction $\varepsilon^\star_{clay}$ as well as the real and
constant permittivity of quartz grain $\varepsilon_{sand}$ and air (\cite{Shen85}, \cite{Olhe81},
\cite{Saar98}, \cite{Sihv00}, \cite{Ishi00}, \cite{Ishi03}, \cite{Cose04}). The dielectric
relaxation behaviour of each process is described by a fractional relaxation model according to
(\ref{eq:relaxation_model}) considering relaxation time distributions $H(\tau)$. This allows the
complete spectrum to fit as a function of water content $\theta_g$ and bulk density $\varrho$ at
constant temperature and pressure with the use of a generalized dielectric response (GDR, Fig.
\ref{fig:GDRFit} and \ref{fig:complex_permittivity_bentonite}):
\begin{equation}\label{eq:relaxation_model_mixture}
\varepsilon_{\mbox{eff}}^\star(\omega)-\varepsilon_\infty=\sum\limits_{i = 1}^3 {\frac{{\Delta
\varepsilon _i}}{{\left( {j\omega \tau _i} \right)^{\alpha _i} + \left( {j\omega \tau _i }
\right)^{\beta _i} }}} - j\frac{{\sigma _{DC}}}{{\omega \varepsilon _0 }}
\end{equation}

\begin{table}\label{tab:fit_data}
  \caption{Parameters of the three relaxation processes from GDR-fitting ($i=[\alpha, \alpha', \beta]$);
  gravimetric water content $w$, bulk density $\varrho$,  porosity $n$, water saturation $S_w$
  and volumetric water content $\theta=S_w n$ of SB-50/50.}
\begin{indented}
 \item[] \begin{tabular}{|l|l|l|l|l|l|l|l||l|}
  \hline
  & 50-0 & 50-1 &  50-2 &  50-3 &  50-4 &  50-5 &  50-6 & Mikrosil \\\hline
  \hline
 $w$ $[\%]$  & 0.6 & 5 & 11.4 & 18.1 & 28.1 & 33.5 & 41.5 & 25.47 \\
 $\varrho$ $[g/cm^3]$  & 1.35 & 1.14 & 1.37 & 1.34 & 1.75 & 1.52 & 1.66 & 1.55 \\
 $n$  & 0.5 & 0.6 & 0.55 & 0.59 & 0.53 & 0.62 & 0.64 & 0.56 \\
 $S_w$ & 0.02 & 0.09 & 0.25 & 0.34 & 0.66 & 0.54 & 0.63 & 0.71 \\
 $\theta$ $[\%]$ & 0.81 & 5.42 & 13.84 & 19.86 & 35.36 & 33.86 & 40.3 & 39.48\\
 \hline
 $\varepsilon_\infty$  & 1.03 & 1.15 & 1.23 & 2.07 & 3.38 & 3.77 & 6.69 & 1.26\\
 $\Delta\varepsilon_\alpha$  & 0.86 & 0.05 & 0.01 & 0.13 & 5.44 & 3.67 & 19.57 & 16.73 \\
 $\tau_\alpha$ $[ps]$& 4.99 & 9.18 & 8.44 & 7.43 & 6.06 & 8.74 & 9.8 & 9.25 \\
 1-$\beta_\alpha$  &0.004 & 0.004 & 0.07 & 0.025 & 0.075 & 0.091 & 0.095 & 0.002 \\
 $\alpha_\alpha$ (fixed)&  0 & 0 & 0 & 0 & 0 & 0 & 0 & 0\\
 $\Delta\varepsilon_{\alpha'}$  & 0.25 & 2.16 & 3.18 & 5.06 & 10.31 & 19.7 & 5.77 & 2.09 \\
 $\tau_{\alpha'}$ $[ps]$& 5.37 & 9.64 & 9.34 & 9.63 & 9.93 & 9.7 & 9.32 & 0.27 \\
 1-$\beta_{\alpha'}$  & 0.004 & 0.081 & 0.081 & 0.075 & 0.003 & 0.082 & 0.034 & 0.048 \\
 $\alpha_{\alpha'}$ (fixed) & 0 & 0 & 0 & 0 & 0 & 0 & 0 & 0 \\
 $\Delta\varepsilon_\beta$  & 0.05 & 1.57 & 4.9 & 11.06 & 43.35 & 62.17 & 74.99 & 0.87 \\
 $\tau_\beta$ $[ns]$& 98.46 & 0.5 & 0.6 & 0.51 & 0.52 & 0.65 & 0.83 & 22.6\\
 1-$\beta_\beta$  & 0.65 & 0.99 & 0.56 & 0.69 & 0.66 & 0.68 & 0.74 & 0.92 \\
 $\alpha_\beta$ (fixed)& 1 & 1 & 1 & 1 & 1 & 1 & 1 & 1\\
 \hline
 $\sigma_{DC}$ $[mS/cm]$ & 6.68E-5 & 0.03 & 0.05 & 0.75 & 3.37 & 4.62 & 5.31 & 0.12 \\\hline
 \end{tabular}
  \end{indented}
\end{table}

A Shuffled Complex Evolution Metropolis (SCEM-UA) algorithm is used to find best GDR fitting
parameters \cite{Heimo04}, Tab. \ref{tab:fit_data}). This algorithm is an adaptive evolutionary
Monte Carlo Markov Chain method inspired by the SCE-UA global optimization algorithm of
\cite{Duan92} and combines the strengths of the Metropolis algorithm \cite{Metr53}, controlled
random search \cite{Pric87}, competitive evolution \cite{Holl75}, and complex shuffling
\cite{Duan92} to obtain an efficient estimate of the most optimal parameter set, and its
underlying posterior distribution, within a single optimization run. The resulting relative error
of each parameter is less than 3\%.
\begin{figure}[ht]
\begin{indented}
 \item[]
  \includegraphics[scale=2.48]{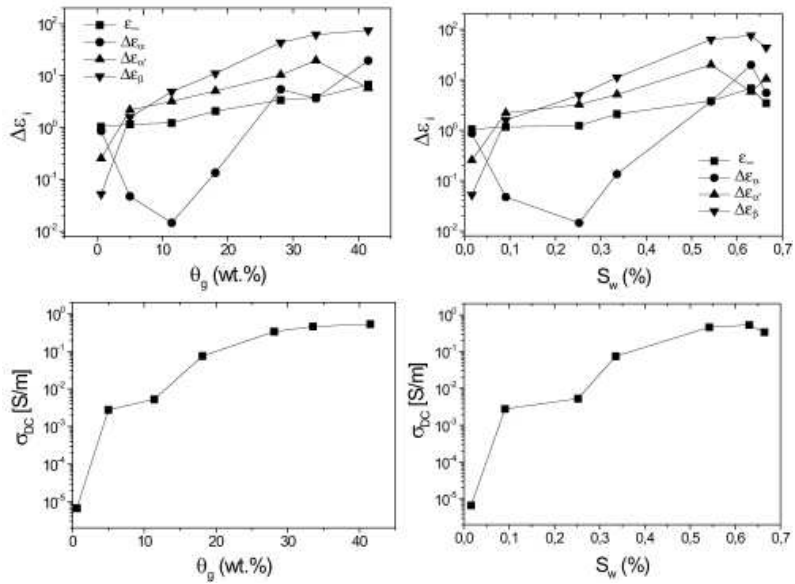}
  \caption{(top) Relaxation strength $\Delta\varepsilon_i$ of the i-th process in comparison to the
  relative high frequency permittivity $\varepsilon_\infty$ as a function (left) of gravimetric water content and
  (right) water saturation $S_w$. (bottom) Apparent direct current electrical conductivity $\sigma_{DC}$.}
  \label{fig:relaxation_parameter_bentonite}
\end{indented}
\end{figure}

The relaxation parameters obtained from GDR-fit are presented in Fig.
\ref{fig:relaxation_parameter_bentonite}. Relaxation strength $\Delta\varepsilon_i$ of each
process and relative high frequency permittivity $\varepsilon_\infty$ depend on moisture content.
Above a gravimetric water content of $\approx 30 wt.\%$ the $\alpha'$-process decrease and the
$\alpha$-process strongly increase. This suggest an increase of the primary $\alpha$-relaxation at
the expense of the bound water process $\alpha'$. In contrast at the highest saturation level of
0.66 at a porosity of 0.53 primary $\alpha$ and low frequency $\beta$-process decrease. Relaxation
time $\tau_i$ and distribution parameter $\beta_i$ are nearly constant. An exception represents
sample 50-0 which shows clear deviation from the general trend but because of the very low
gravimetric water content of 0.6 wt\%. Without swelling an increasing saturation reduces the
gas-filled  pore space, while the effective pore space available for the fluid becomes larger.
This leads to the simple calibration models mentioned in the introduction to determine the
volumetric water content with dielectric measurements. Due to the swelling of the clay minerals in
the process of the hydration the distribution of immobile and effective pore space cannot be
considered as constant. A circumstance which complicate a careful realization of accurate
measurements is the fact that a multitude of variables of the measurement conditions and the
sample preparation essential determine the intensity of the swelling and the resulting change of
the pore structure (\cite{Agus05}).

The results show the potential of the chosen approach but a detailed explanation of this complex
behavior is beyond the scope of this paper. In general, there is a need of further systematic
investigations by broadband dielectric spectroscopy of saturated and unsaturated soils under
consideration of the swelling process and with an utilisation of microscopic modelling.

\begin{figure}[b]
\begin{indented}
 \item[]
  \includegraphics[scale=0.70]{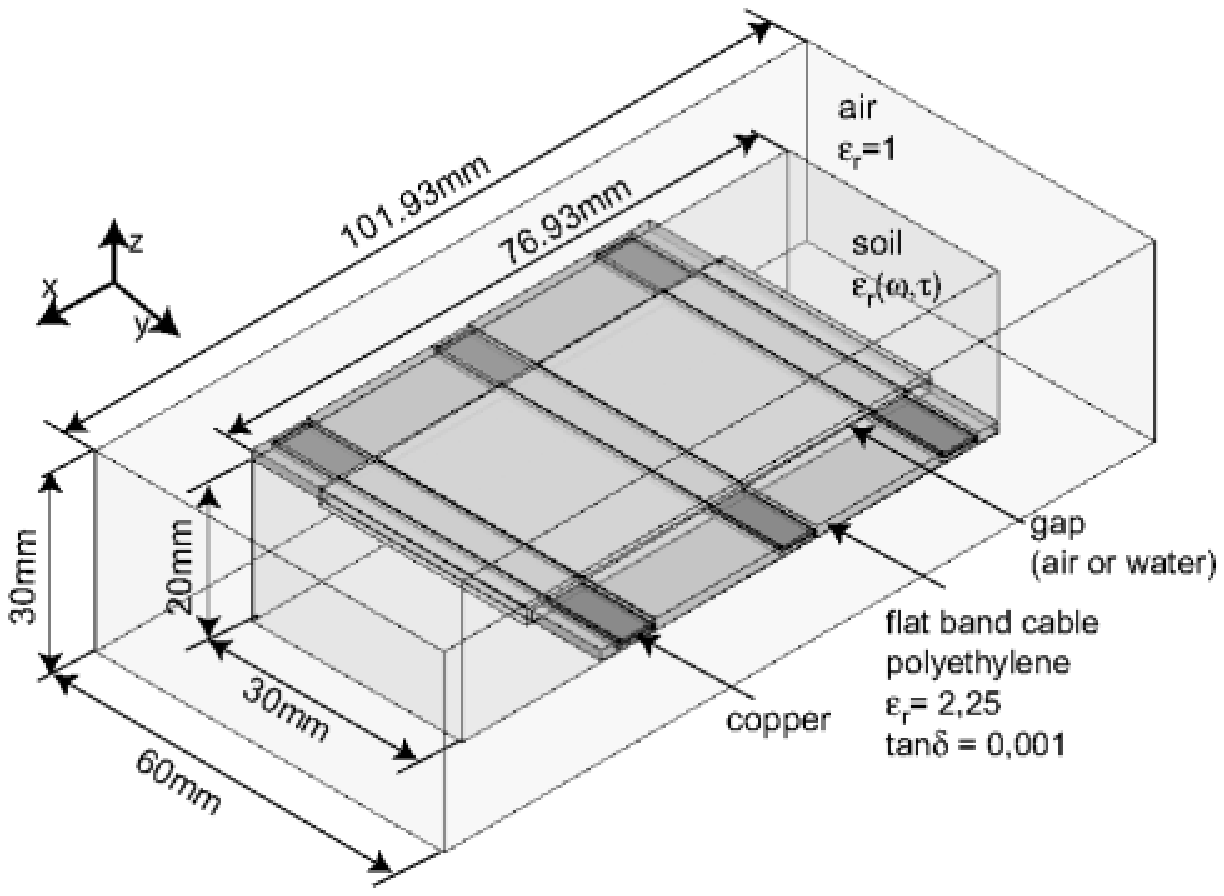}
  \includegraphics[scale=1.40]{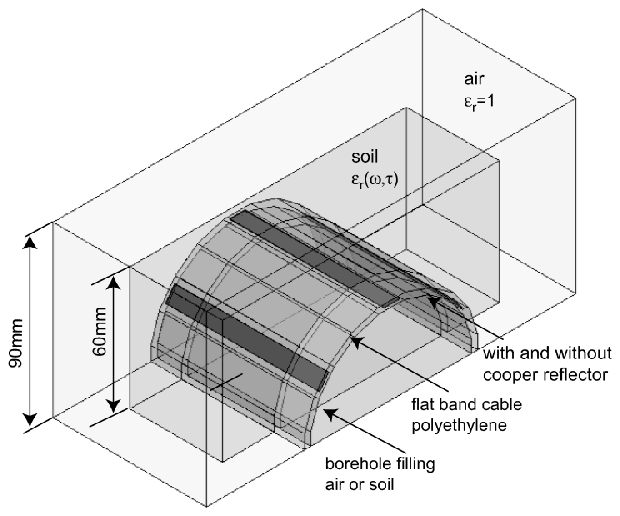}
  \caption{Model geometry of (top) flat band cable surrounded by saturated and unsaturated
  soil with a gap filled with air or water and (bottom) cable sensor pressed at a borehole-wall with air or soil as borehole filling.}\label{fig:geometry_HFSS}
\end{indented}
\end{figure}

\section{HFSS-Simulations}
The transfer or scattering function $S_{ij}(\omega)$ of the flat band cable section (Fig.
\ref{fig:geometry_HFSS}) was simulated by finite element modelling (commercial software from
Ansoft: High Frequency Structure Simulator-HFSS\footnote{HFSS is a standard simulation package for
electromagnetic design and optimization.}) under certain conditions: (i) in direct contact to the
surrounding material (air, water of various salinities, different synthetic and natural soils
(sand-silt-clay mixtures)), (ii) with consideration of a defined gap of various size (total high 2
mm, 3 mm, 5 mm, 7 mm or 10 mm) filled with air or distilled water and (iii) cable sensor pressed at
a borehole-wall.

In HFSS tangential element basis function interpolates field values from both nodal values at
vertices and on edges. Surfaces of the structure (air-box) in the y-z an x-y plane are radiation
boundaries and the second-order radiation boundary condition is used
\begin{equation}\label{eq:Radiation}
\nabla\times\vec{E}_t=jk_0\vec{E}_t-\frac{j}{k_0}\nabla_t\times(\nabla_t\times\vec{E}_t)+
\frac{j}{k_0}\nabla_t(\nabla_t\bullet\vec{E}_t)
\end{equation}
where $\vec{E}_t$ is the component of the E-field that is tangential to the surface and $k_0$ is
the free space phase constant. The second-order radiation boundary condition is an approximation of
free space. The accuracy of the approximation depends on the distance between the boundary and the
object from which the radiation emanates. For this reason a sensitivity analysis was carried out
for a total height of the structure between 20 mm and 50 mm in dependence of the material
properties. The influence of the boundary layer for the simulation into air can be neglected for a
distance from the cable sensor greater than 12 mm so a minimum height of 25 mm is used.

Surfaces in the x-z plane are wave ports. HFSS assumes that each wave port is connected to a
semi-infinitely long waveguide that has the same cross-section and material properties as the
port. When solving for the S-parameters, HFSS assumes that the structure is excited by the natural
field patterns (modes) associated with these cross-sections. The 2D field solutions generated for
each wave port serve as boundary conditions at those ports for the 3D problem. The final field
solution computed must match the 2D field pattern at each port.
\begin{figure}[ht]
\begin{indented}
 \item[]
  \includegraphics[scale=1]{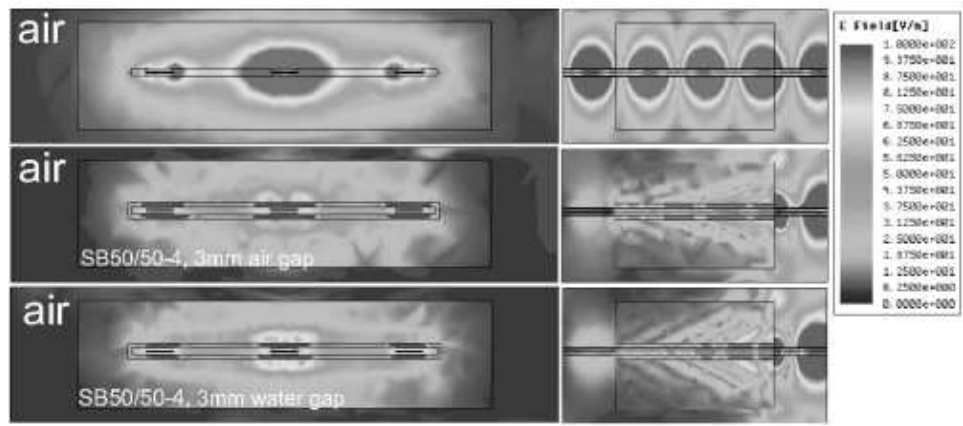}
  \caption{Electric field distribution $@12.5GHz$ for the investigated flat band cable surrounded by air,
  sand-bentonite-mixture (SB 50/50-4) with $\theta_g=$28.14wt.\% and $\rho=1.79g/cm^3$ as well as a defined 3mm air or water gap.
  (left) Cross section, (right) longitudinal section of middle conductor.}\label{fig:gap_fields_HFSS}
\end{indented}
\end{figure}

The simulation is performed with a $\lambda/3$ based adaptive mesh refinement at solution frequency
of 1 MHz, 10 MHz, 0.1 GHz, 1 GHz and 12.5 GHz. Broadband complex S-Parameter are calculated with an
interpolating sweep in frequency range 1 MHz - 12.5 GHz with extrapolation to DC. The
electromagnetic field distribution, S-parameter and step response (200 ps rise time) of the
structure were computed in reflection and transmission mode.
\begin{figure}[ht]
\begin{indented}
 \item[]
  \includegraphics[scale=1.55]{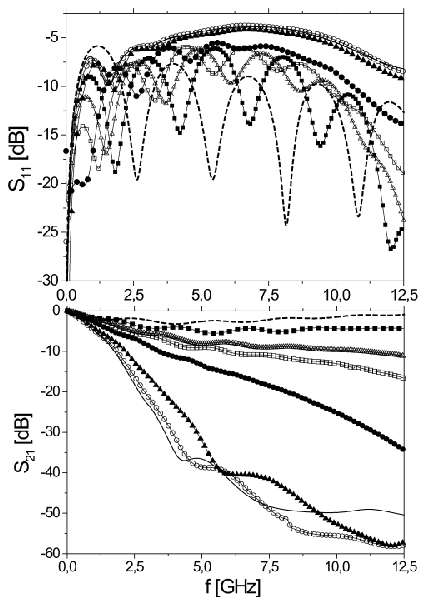}
  \includegraphics[scale=1.35]{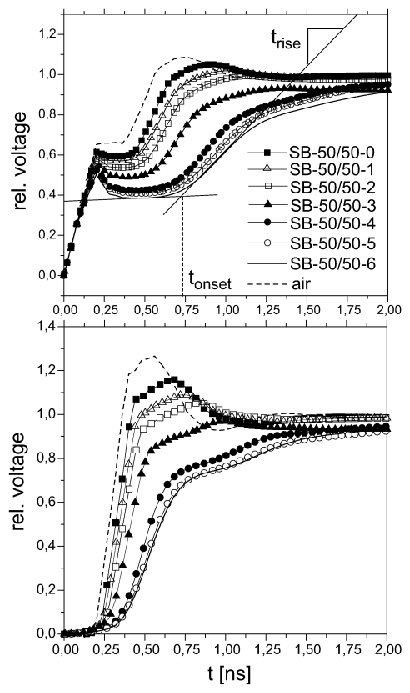}
  \caption{(left) Input return loss magnitude or reflection coefficient $S_{11}$ and forward
  transmission or transmission coefficient $S_{21}$ as well as (right) TDR-waveform in reflection
  and transmission mode for simulated flat band cable structure, surrounded by air and sand bentonite
  mixture of various water contents and bulk densities (see Tab. \ref{tab:fit_data}).}\label{fig:HFSS_results}
\end{indented}
\end{figure}

\section{Discussion}

The simulation adequately reproduces the spatial and temporal electrical and magnetic field
distribution in comparison with 2D-FE investigations of \cite{Hueb05} (Fig.
\ref{fig:gap_fields_HFSS}). Fig. \ref{fig:HFSS_results} represents reflection and transmission
factor as well as the corresponding TDR waveform for the sand-bentonite mixture at various water
contents and bulk densities in reflection and transmission mode. As a reference material the
simulation results for the cable sensor surrounded by air are included. As expected, the
appropriate resonances in the reflection coefficient $S_{11}$ shift with rising water content to
deeper frequencies and the attenuation increases. In time domain onset travel-time as well as rise
time of the TDR signals are analyzed. As is clearly recognizable in both reflection and
transmission mode, onset time increase and rise time decrease with rising water content.

Qualitatively the numerical calculation shows that sensitivity characteristic of the cable sensor
changes along the sensor in dependents of the dielectric relaxation behavior of the surrounding
material (Fig. \ref{fig:gap_fields_HFSS}). The investigated high-lossy sand-bentonite mixture cause
in dependence of increasing gravimetric water content $\theta_g$ and bulk density $\varrho$ an
increase of TDR signal rise time as well as a strong absorption of multiple reflections. This leads
to a frequency dependent decrease of spatial resolution and penetration depth (sensitivity region
around the sensor) along the flat band cable and fixed maximal length of the moisture-sensor
available for application.

\begin{figure}[ht]
\begin{indented}
 \item[]
  \includegraphics[scale=3]{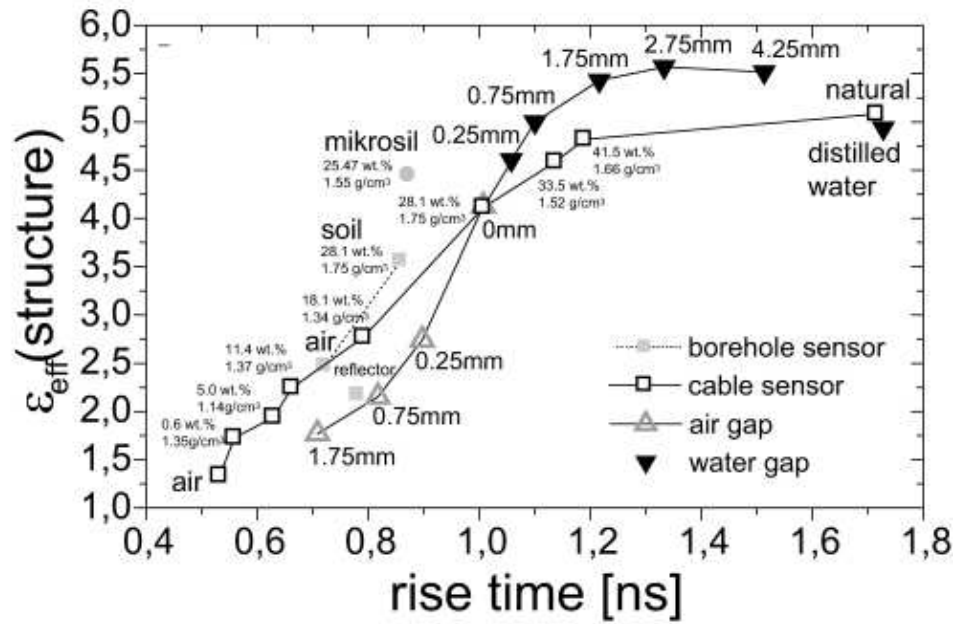}
  \caption{Real effective relative permittivity $\varepsilon_{\mbox{eff}
  }=2l/t_{\mbox{onset}}$  according to equation (\ref{eq:epsilon_trawel_time}) plotted
  against TDR rise time $t_{\hbox{rise}}$ (in reflection mode) for all sensor configurations and investigated cases.}\label{fig:eps_rise_plot}
\end{indented}
\end{figure}

Coupling problems caused by air or water gaps lead to dramatic travel time distortion even for
very small gaps (Fig. \ref{fig:eps_rise_plot}). The air filled gap with a thickness of 0.25mm on
both sides of the cable sensor already leads to the drastic underestimation of water content of
36\%. In contrast a drastic overestimation occurs in the case of a water filled gap for the same
gap size. Further increasing gap size leads to a maximum in the characteristic (anomal behavior)
at a gap thicknesses of approximately 2.7 mm, whereby up to the complete filling of the structure
with water the effective dielectric constant decrease and the rise time increase. An explanation
of this effect and the consequences linked with it are however still pending and require further
numerical and experimental investigations. Moreover, with HFSS it is not possible to consider an
exchange of material between air or water-filled gap and surrounding medium.

As a consequence the gap work as quasi waveguide, i.e. the influence of the surrounding medium is
strongly reduced but changes in dielectric properties along the cable sensor are reproduced.
\section{Conclusion}

In this study, the spatial sensor characteristics of a 6cm TDR flat band cable section is simulated
with finite element modelling in combination with dielectric spectroscopy.

For this purpose the dielectric relaxation behaviour of saturated and unsaturated soils is examined
in the frequency range 50MHz-20GHz. The dielectric relaxation behaviour is described with the use
of a generalized fractional relaxation model. With this approach the frequency depended dielectric
permittivity is determined based on a parametrization of each relaxation processes as a function of
water content and porosity. This enables a development of improved calibration strategies. However,
there is a need of further systematic investigations by broadband dielectric spectroscopy of
saturated and unsaturated soils under consideration of the swelling process and with an utilisation
of microscopic modelling.

The three-dimensional numeric finite element simulation in HFSS provide an informative basis of the
sensor characteristics under consideration of the frequency dependence of the measured complex
dielectric permittivity. It is shown, that an air or water gap between sensor and soil leads to
dramatic under or overestimation of water content already for a gap thickness of 0.25 mm upper and
below the cable sensor. Therefore, the application of the flat cable as a moisture sensor requires
an accurate installation technique of cable-like elements (especially for long sensors). Moreover,
the spacial sensitivity characteristics of the cable sensor changes along the sensor in dependents
of the dielectric relaxation behavior of the surrounding material.  For this reason, the precise
determination of soil moisture profiles requires an improved TDR-waveform analysis strategy under
consideration of the change of the sensitive area along the cable sensor.

A disadvantage of the used FEM with HFSS is the lack of consideration an exchange of material
between air or water-filled gap and surrounding medium. In addition, to achieve sufficient accuracy
the numerical simulations with HFSS requires a high degree of computational cost in terms of
computational time and memory. Further, theoretical, numerical, and experimental investigations in
conjunction with reconstruction algorithms have to point out to what extent the accuracy of water
content and porosity profiles can be determined in strong dispersive, high lossy materials.

\section*{References}
\bibliographystyle{plain}
\bibliography{Literatur}

\end{document}